\documentclass[preprint,12pt]{elsarticle}

\usepackage{amssymb}
\usepackage{subfig}
\journal{Nuclear Instruments and Methods in Physics Research Section A}

\begin{document}

\begin{frontmatter}

\title{Proton induced Dark Count Rate degradation in 150-nm CMOS Single-Photon Avalanche Diodes}

\author[unina,infn]{M. Campajola \corref{mycorrespondingauthor}}
\ead{macampajola@na.infn.it}

\author[unina,infn]{F. Di Capua, \corref{mycorrespondingauthor}}
\ead{dicapua@na.infn.it}

\cortext[mycorrespondingauthor]{Corresponding author}

\author[dimes]{D. Fiore}
\author[infn,cnr]{E. Sarnelli}
\author[unina,infn]{A. Aloisio}

\address[unina]{Universit\'a di Napoli 'Federico II', Dipartimento di Fisica 'E. Pancini', Napoli, Italy}
\address[infn]{INFN, Sezione di Napoli, Napoli, Italy}
\address[dimes]{Universit\'a della Calabria, Dipartimento DIMES,
Cosenza, Italy}
\address[cnr]{CNR-SPIN, Napoli, Italy}

\begin{abstract}
Proton irradiation effects on a Single-Photon Avalanche Diodes (SPADs) device manufactured using a 150-nm CMOS process are presented. An irradiation campaign has been carried out with protons of 20 MeV and 24 MeV on several samples of a test chip containing SPADs arrays with two different junction layouts. The dark count rate distributions have been analyzed as a function of the displacement damage dose. 
Annealing and cooling have been investigated as possible damage mitigation approaches. We also discuss, through a space radiation simulation, the suitability of such devices on several space mission case-studies.

\end{abstract}


\end{frontmatter}


\section{Introduction}
Single-Photon Avalanche Diodes (SPADs) are playing a significant role in the development of high-performance detectors for the new era High Energy Physics (HEP) and space experiments. 
They are semiconductor photon detectors operating in Geiger-mode, i.e., biased above the diode breakdown voltage. Due to their working principle, SPADs are capable of a substantial internal gain (G$\sim10^6$) with no need for pre-amplification, resulting in single-photon sensitivity. Furthermore, they can provide excellent timing and spatial resolution, reaching a few tens of picoseconds and micrometers, respectively \cite{Cova}. 
In the early 2000s, SPAD-based pixel arrays realized in CMOS technology have been demonstrated \cite{Rochas}. The use of CMOS technology for the design and the fabrication of SPADs led to the monolithic integration of additional pixel circuitry for avalanche quenching and signal processing, like Counter, Time to Digital Converter, etc. This resulted in a significant advantage in terms of signal temporal response, detector miniaturization and portability, low power consumption, and low fabrication cost.
Due to all these extraordinary features, CMOS SPADs are a promising alternative to traditional photon detectors \cite{Palubiak}. 

Many applications in the field of HEP are under investigation, such as vertex detectors for charged particles \cite{Vilella,Apix}, aerogel RICH detectors \cite{farich}, ToF detector \cite{tof}, etc.
Even in many space applications, CMOS SPADs are very promising detectors, especially considering their low power dissipation and the possibility of compactness. 
In particular, SPADs are an appealing alternative to classical PMTs in scintillation light detectors \cite{dsipm, jinstspad} for terrestrial and space cosmic-ray experiments. Even LIDAR, Rendez-vous, Time of Flight 2D reconstruction are possible fields of application.

In spite of all their advantages, CMOS SPADs suffer from a high level of spurious pulses whose mean rate is indicated as Dark Count Rate (DCR). One of the main mechanism responsible for dark counts in SPADs is the thermal generation of free carriers within the depletion region. It is closely related to the presence of impurities and crystal defects that introduce localized energy levels near the middle of the band-gap. These act as efficient electrons and holes generation-recombination (G-R) centers according to the Shockley-Read-Hall (SRH) model \cite{Shockley, Hall}. Beyond thermal effects, another major contributor to DCR is band-to-band or trap-assisted tunneling \cite{Kane, Xu}, enhanced by the high and shallow doping profiles typical of CMOS implementation. 

Defects and impurities can be introduced in the detector not only in the manufacturing process, but also during its use, i.e., if operating in radiation-full environments. It is well known that exposure to particle flux can result in the production of bulk and insulator defects \cite{Srour}. Therefore, a sufficient radiation hardness is a mandatory requirement for CMOS SPADs in order to withstand many years of operation in a radiation environment. 
Many papers in literature report on radiation effects in custom SPADs or in SiPM \cite{Li, Andreotti, Moscatelli}. Some recent works focused on radiation DCR induced effects in CMOS SPADS \cite{Charbon, Charbon2, Ratti, campajola_2019, philips} while others focused on Random Telegraph Signal effects \cite{Karami, Karami2, DiCapua}, but  state of the art still need further insights.  

This work aims to investigate the degradation induced by protons on a given electrical parameter, i.e., the dark count rate of a monolithic SPADs detector fabricated in a 150 nm CMOS process. 
For this purpose, an irradiation campaign has been carried out with protons of 20 MeV and 24 MeV to induce displacement damage effects on several samples of a test chip containing SPADs arrays with two different junction layouts. The DCR has been measured as a function of the dose on a large amount of SPADs, showing that, for this kind of devices, displacement damage could be a serious issue.
Furthermore, we addressed some interesting results on the radiation-induced DCR increase origin and some possible mitigation approaches.

The obtained results have been investigated in the framework of several space mission case-studies. Expected radiation levels have been estimated by means of the web-tool software SPENVIS \cite{Spenvis}, simulating several space mission orbits, with different inclinations and shield thickness.

\section{Experimental}
\subsection{Device layout}
\begin{figure}
\centering
\subfloat{%
  \includegraphics[width=8.5cm]{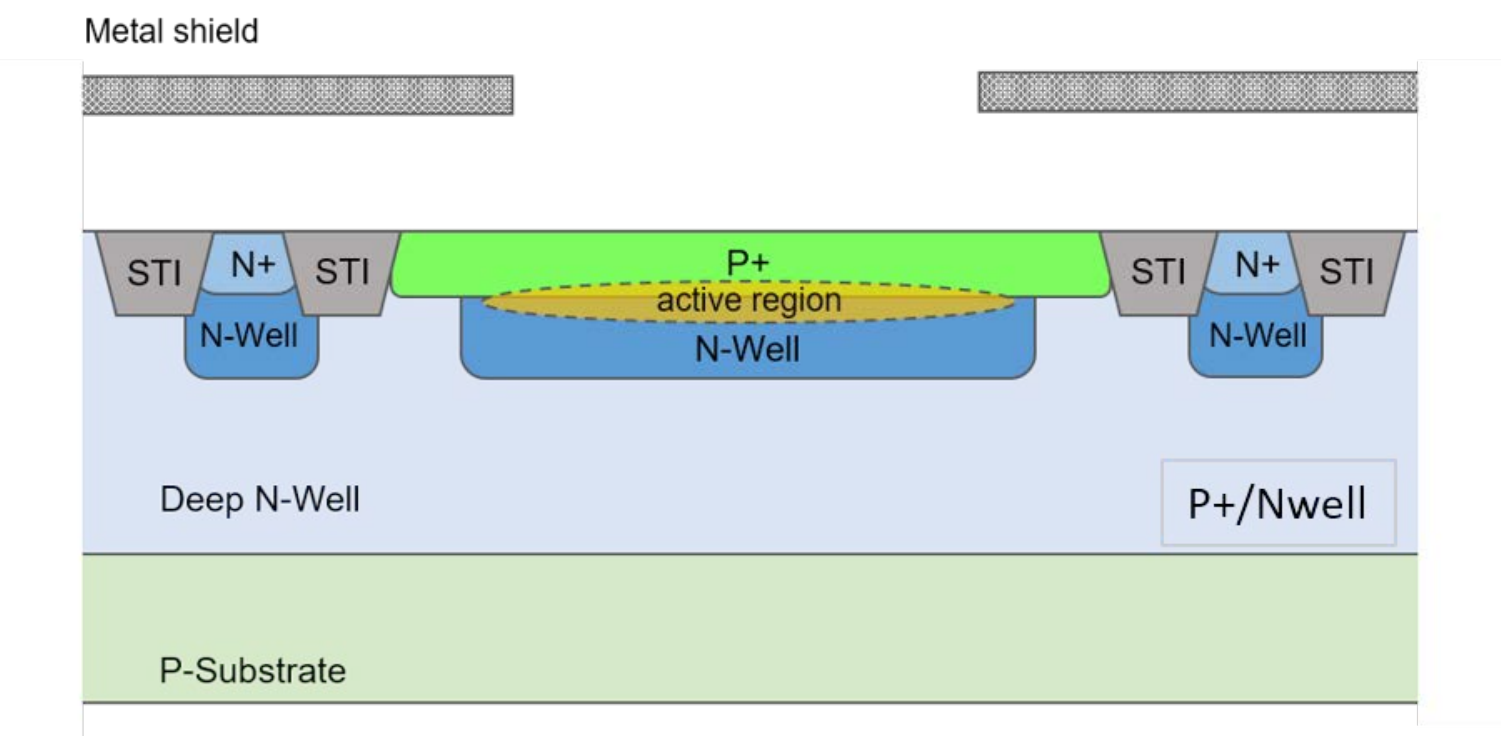}%
  }
  
\subfloat{%
  \includegraphics[width=8.5cm]{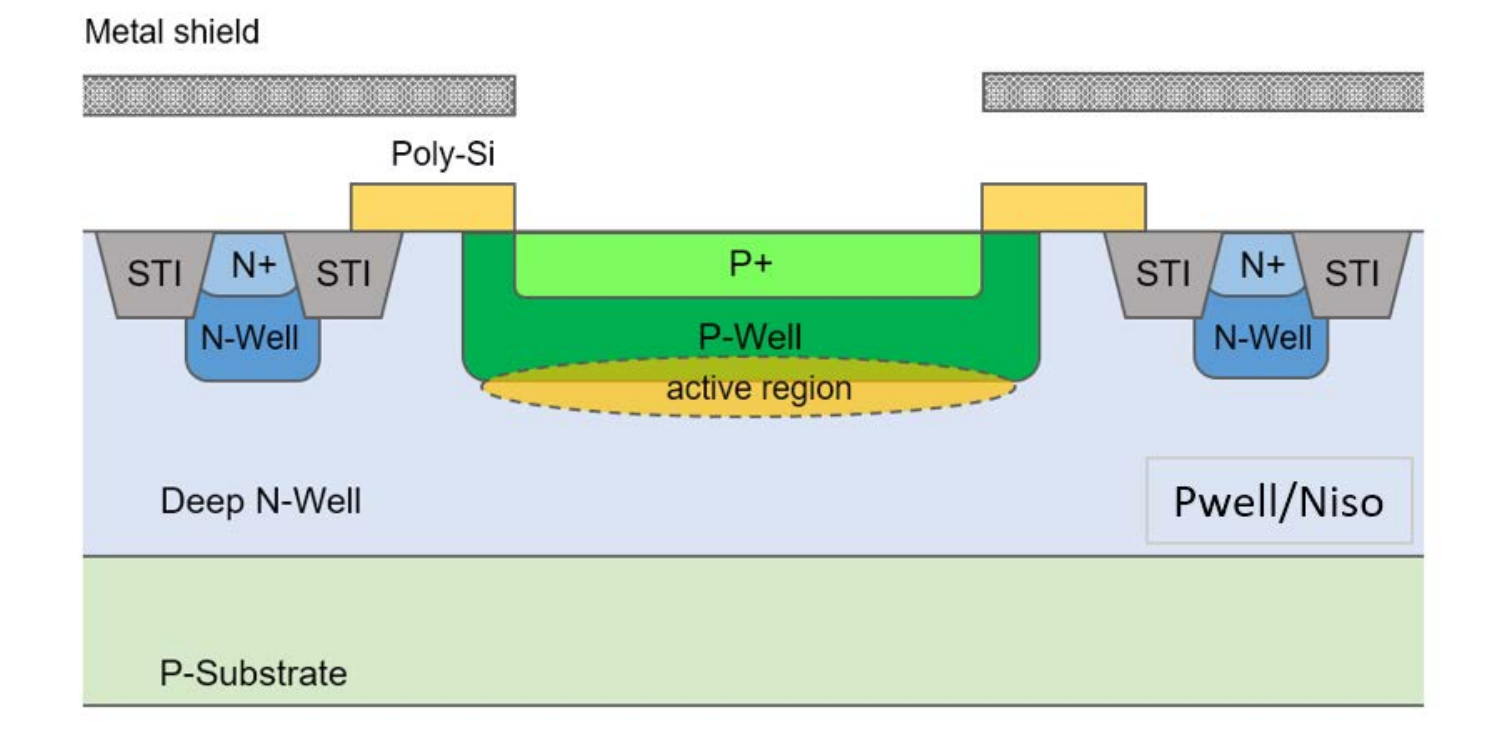}%
  }

\caption{Cross section of SPAD structures: (top) P+/Nwell and (bottom) Pwell/Niso layouts.}\label{SPAD_layout}

\end{figure}

The devices investigated in this work are test chips provided by Fondazione Bruno Kessler (FBK) \cite{FBK}, containing several architectures of SPADs. 
A first structure is based on a P+/Nwell junction (Fig. \ref{SPAD_layout}, top). The guard ring, essential in order to avoid premature periphery edge breakdown, is obtained by blocking both P-well and N-well at the borders of the junction with a deep N-Well implantation. A second SPAD structure is based on a Pwell/Niso junction  (Fig. \ref{SPAD_layout}, bottom). In this case, the guard ring is obtained by blocking both P-well and N-well at the junction periphery. A poly-Si gate blocks P+ implantation avoiding it to reach the Shallow-Trench-Isolation (STI) region.

\begin{figure}
\centering
  \includegraphics[width=7cm]{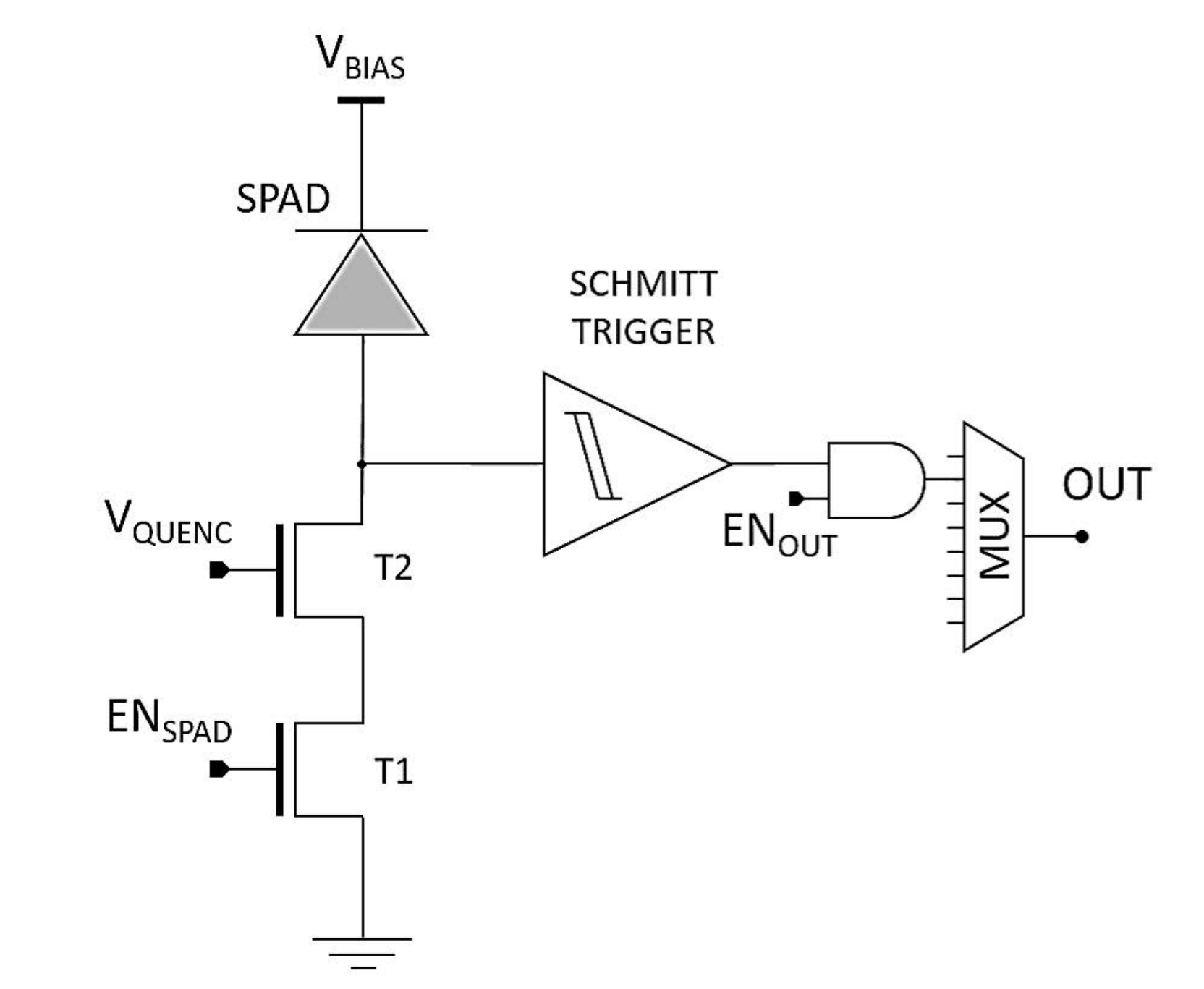}
  \caption{Schematic diagram of the SPAD front-end.}\label{SPAD_frontend}
\end{figure}

Each SPAD is integrated with its relative front-end pixel circuit (Fig. \ref{SPAD_frontend}). The SPAD is connected to a quenching transistor T2 which acts as a resistor. An enable transistor T1, in series with T2, is used to pull the SPAD below the breakdown voltage and to shut it down. 
A Schmitt-trigger comparator converts the voltage pulse from the SPAD to a digital output signal. Each SPAD can be individually enabled trough row and column decoder and a multiplexer (MUX) is used to connect one SPAD at a time to the output. 

The test chip contains SPADs with four different optical window sizes: 5, 10, 15, 20 $\mu m$. Each one is arranged in linear arrays of 20 SPADs and square matrices of 5x5 SPADs. In this paper, we focused on the 10x10 $\mu m^2$ SPADs, arranged  in both arrays and matrices.
More details about the chip design are given in \cite{Pancheri,Hesong}.

\subsection{Measurement setup}
\begin{figure}[h!]
\centering
  \includegraphics[width=8.5cm]{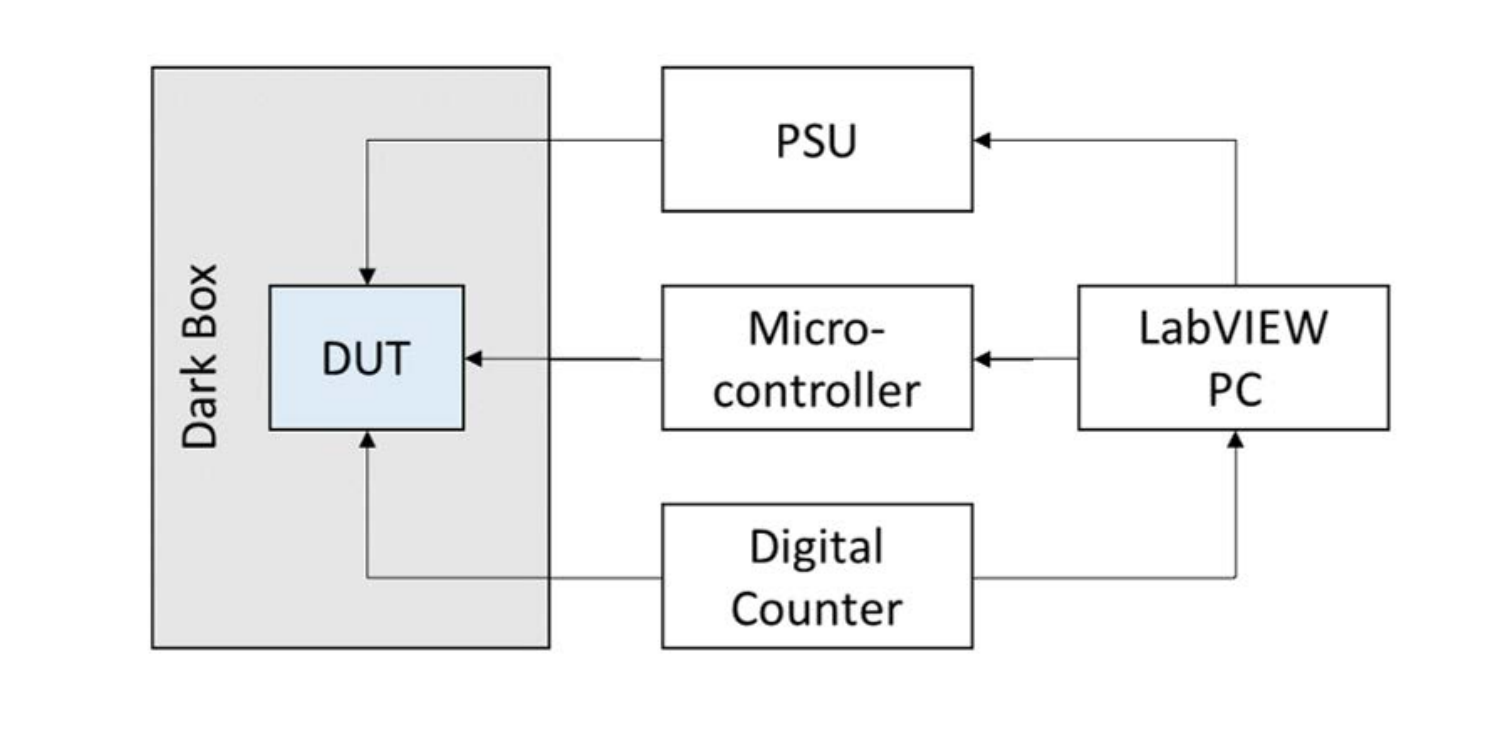}
  \caption{Schematic diagram of experimental setup.}\label{setup}
\end{figure}

A schematic layout of the experimental setup is shown in Fig. \ref{setup}. A dedicated circuit board provides both power supply to the read-out circuit on-chip and the SPAD bias. The chip output signal is sent to an oscilloscope and to a digital counter. In order to enable a certain pixel and connect it to the chip output, a  digital serial pattern is sent to the MUX by means of an external micro-controller. In order to have a fully automatized measurement procedure, SPAD power supply, digital counter, and micro-controller have been connected through a serial bus to a personal computer and interfaced by means of a LabVIEW software.

\section{Pre-irradiation characterization}
\begin{figure}
\centering
  \includegraphics[width=8.5cm]{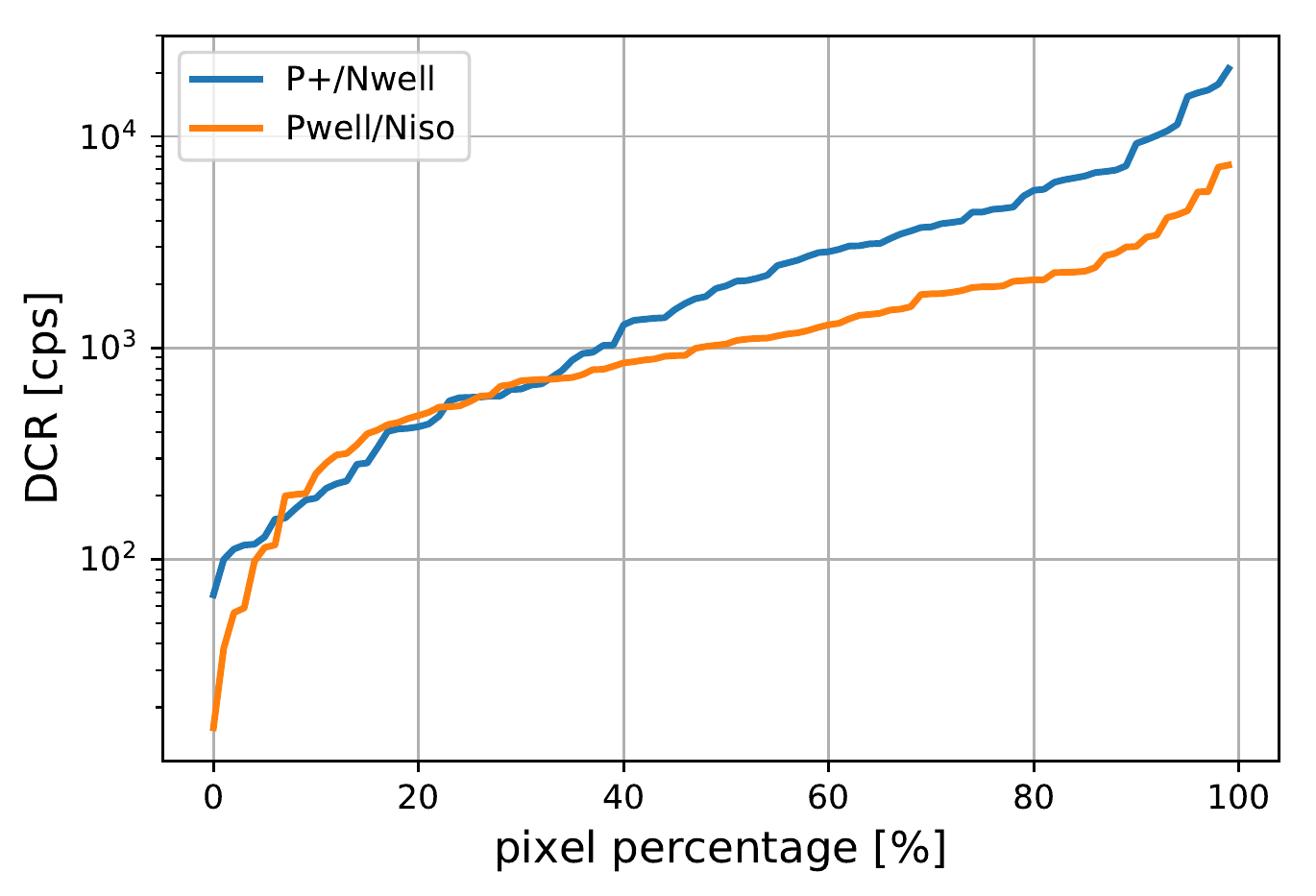}
  \caption{Cumulative distribution of the DCR for the two SPADs structures.}\label{cumulative_before}
\end{figure}

In this section, the DCR characterization of the Devices Under Test (DUTs) before irradiation is showed. DCR measurements have been performed in a dark environment at a controlled temperature of 25 $^\circ$C. 
After measuring the breakdown voltage  from the DCR-voltage characteristics, we obtained the DCR at  3.3V over bias voltage.
Table \ref{tab:dcr} summarizes the DCR characterization for the two SPADs structures. These results include measurements on all the chips that were irradiated later. It can be observed that P+/Nwell layout exhibits higher DCR values than the Pwell/Niso structures (Fig. \ref{cumulative_before}). This behaviour can be addressed to the electric field shape, more peaked in P+/Nwell type \cite{pancheri_field}.

\begin{table}
    \caption{DCR characterization before irradiation.}
    \label{tab:dcr}
    \centering
    \begin{tabular}{lcc}
        \hline
        \hline
        Type & Median  & Mean  \\
         & DCR [cps] & DCR [cps] \\
        \hline
         P+/Nwell  &  1938  &  3342   \\
         Pwell/Niso    &  1036  &  1448  \\
        \hline
        \hline
    \end{tabular}
\end{table}

\section{Proton irradiation test}
A proton irradiation test has been performed on seven CMOS SPAD chips at the Laboratori Nazionali del Sud (LNS, Italy). 

\begin{figure}[t!]
\centering
  \includegraphics[width=8.5cm]{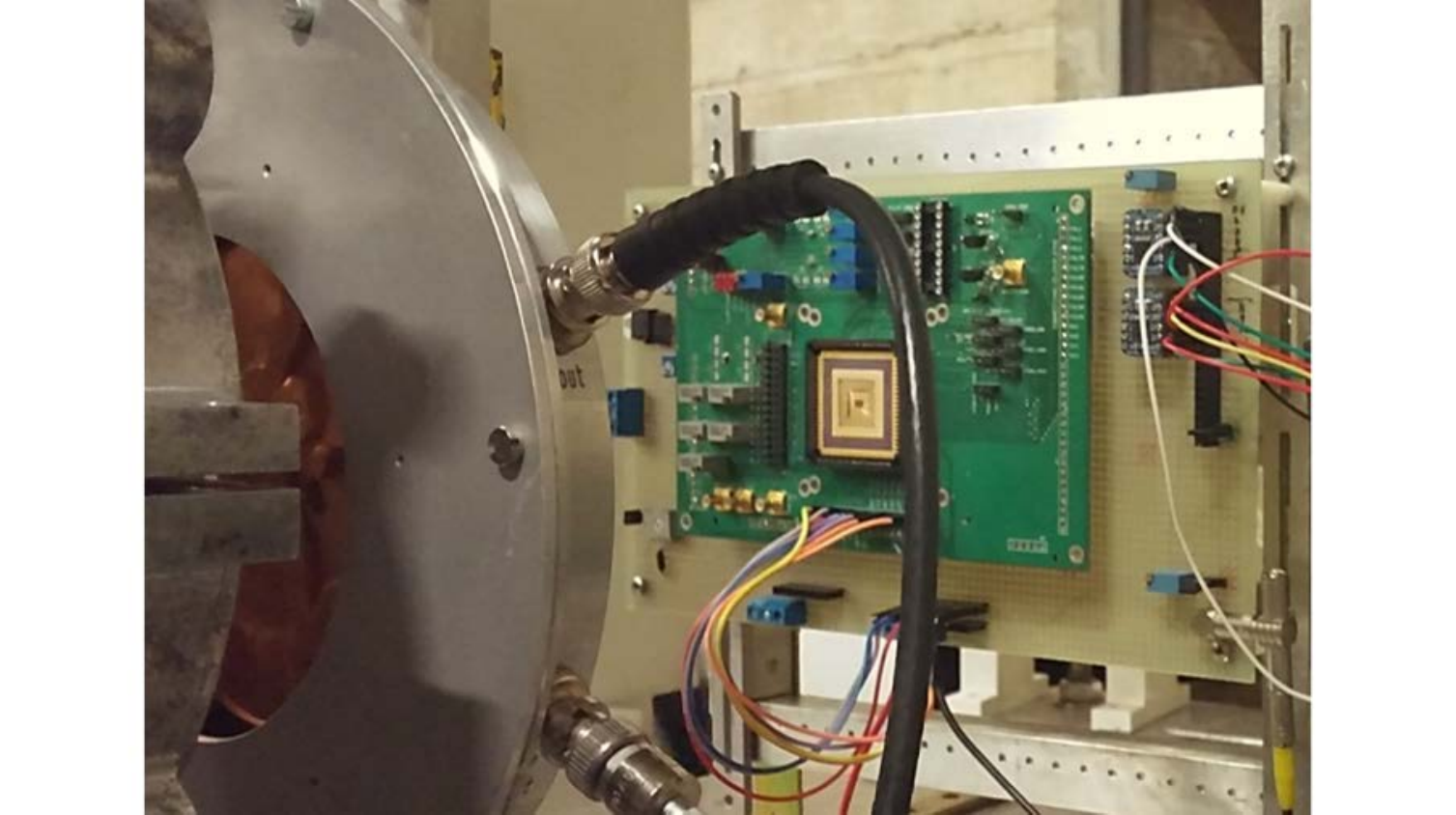}
  \caption{Test beam setup at the Tandem accelerator at Laboratori Nazionali del Sud (INFN-LNS, Italy).}\label{lns}
\end{figure}

Irradiations have been carried out at Tandem accelerator with a 20 MeV and 24 MeV proton beam extracted in air (Fig. \ref{lns}). 
The beam energy on the DUT has been estimated by using both SRIM \cite{srim} and FLUKA \cite{fluka} packages. Taking into account the energy loss along the beam-line and the quartz cover of the chip, the mean beam energy on chip results in 16.4 MeV and 21.1 MeV, respectively. 
The DUT has been carefully aligned with the beamline center by using a laser pointing system.
The beam was collimated through a circular collimator. Its spot diameter ($d$ = 16 mm) and uniformity at the DUT position have been measured by means of a radio-chromic film.
On a region of 10x10 $mm^2$ around the center of the beam spot,  the beam intensity had a variation of $\pm$ 5\% with respect to the average value. This region widely contains, within the alignment error, the tested chip area (2x3 $mm^2$).
The delivered proton charge during each irradiation run has been measured using an ionization chamber, and taking into account the beam profile, the corresponding fluence calculated.
Table \ref{tab:irradiation} summarizes the dosimetry for each irradiation of the chips. 
The Total Ionizing Dose (TID) and the Displacement Damage Dose (DDD) have been respectively calculated as $ \textrm{TID}=\textrm{LET}\cdot \Phi $ and $ \textrm{DDD} = \textrm{NIEL} \cdot \Phi$, where the LET has been obtained by means of SRIM, the NIEL by means of a NIEL calculator  software \cite{NIEL} and $\Phi $ represents the proton fluence. 


\begin{table}[t!]
    \caption{Proton irradiation test summary. }
    \label{tab:irradiation}
    \centering
    \begin{tabular}{ccccc}
        \hline
        \hline
        Chip & Fluence & Energy & TID & DDD \\
            & [p/cm$^2$] & [MeV] & [krad] & [TeV/g] \\
        \hline
        1   & 6.7$\times10^{8}$  & 16.4  & 2.3  & 45   \\
        2   & 1.4$\times10^{9}$ & 16.4  & 4.8  & 94  \\
        3   & 2.0$\times10^{9}$ & 16.4  & 7.1  & 139  \\
        4   & 2.8$\times10^{9}$ & 16.4  & 9.6  & 188  \\
        5   & 4.1$\times10^{9}$ & 16.4  & 14.5 & 283  \\    
        6   & 6.7$\times10^{9}$ & 16.4  & 23.5 & 457  \\
        7   & 9.1$\times10^{10}$ & 21.1  & 30.5 & 608  \\
        \hline
        \hline
    \end{tabular}
\end{table}

\begin{figure}[t!]
\centering

\subfloat{%
  \includegraphics[width=8.5cm]{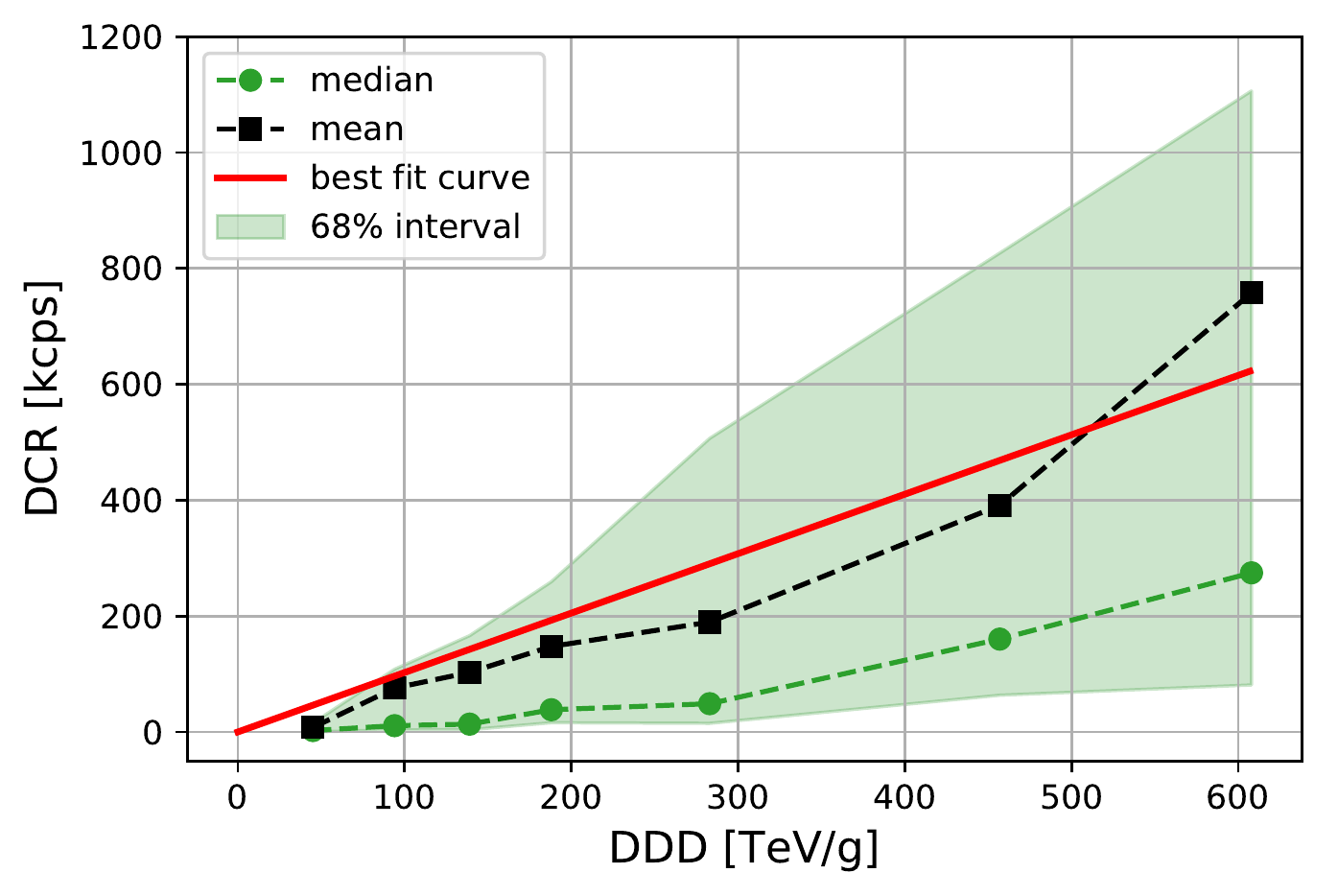}%
  }

\subfloat{%
  \includegraphics[width=8.5cm]{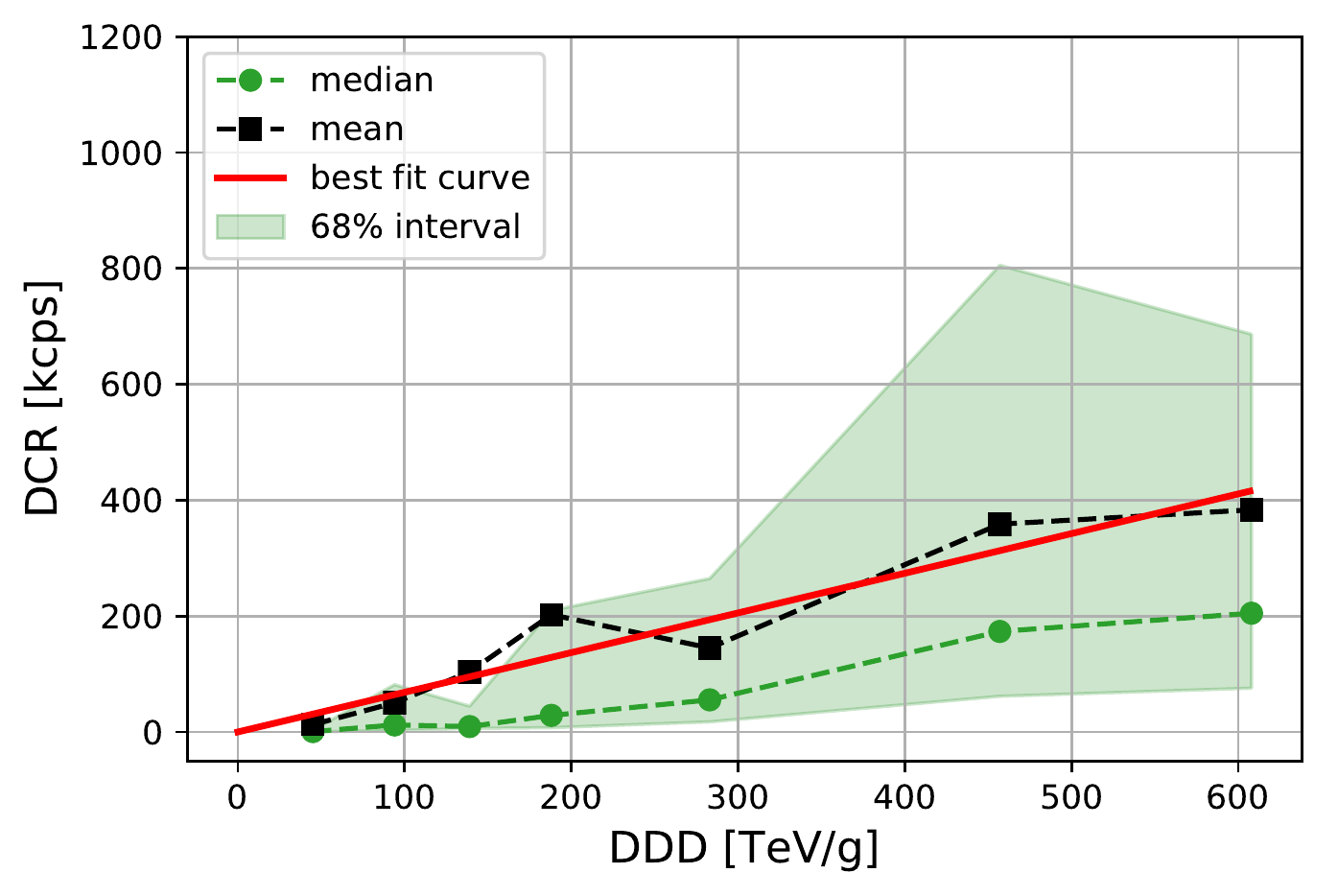}%
  }

\caption{DCR increase as a function of the displacement damage dose. Case of P+/Nwell (top) and Pwell/Niso (bottom) layouts. The mean DCR values have been fitted with a $y=ax$ function: P+/Nwell $a=(1025\pm96)\,\textrm{cps/TeV\,g}$ - Pwell/Niso $a=(685\pm52)\,\textrm{cps/TeV\,g}$. The green band shows the 68\% population interval around the median.}\label{dcrddd}

\end{figure}
During the irradiation, the devices were kept unbiased with all the terminals connected to the ground.
The DCR measurements have been performed after a week from the irradiation. For this study, all the measurements were performed  at 3.3V over bias voltage at room temperature. 
In Fig. \ref{dcrddd}, we report the behaviours of the DCR increase distributions, where the initial DCR was subtracted for every single SPAD, as a function of the delivered displacement damage doses. 


Although data are taken from different samples, it is reasonable to assume that the differences are negligible compared to the expected effects.
It can be observed that the DCR increase is almost linear with respect to the DDD for both P+/Nwell and Pwell/Niso. The trend is very similar between the two structures, except for the higher dose point, where a greater degradation is observed for the P+/Nwell junction.

Regarding the origin of the increase in DCR, several hypotheses can be made.
It is reasonable to assume that the performances of SPADs front-end circuit are not significantly affected by the ionizing and displacement doses delivered in this test.
CMOS transistors under 350-nm node have been demonstrated to be tolerant to ionizing radiation above the Mrad level \cite{trantid}, while no degradation of their operation due to bulk damage is expected \cite{tranddd}.
Similar devices to the ones studied in this work, have been tested with a total ionizing dose irradiation, showing that they are quite immune up to Mrad doses \cite{Ratti}. In our irradiation test, the ionizing dose level is kept below 30 krad, furthermore, the devices were kept off during irradiation. Therefore, the observed degradation effects can be safely ascribed mainly to silicon bulk defects induced by
displacement damage.

\begin{figure}[t]
\centering
  \includegraphics[width=8.5cm]{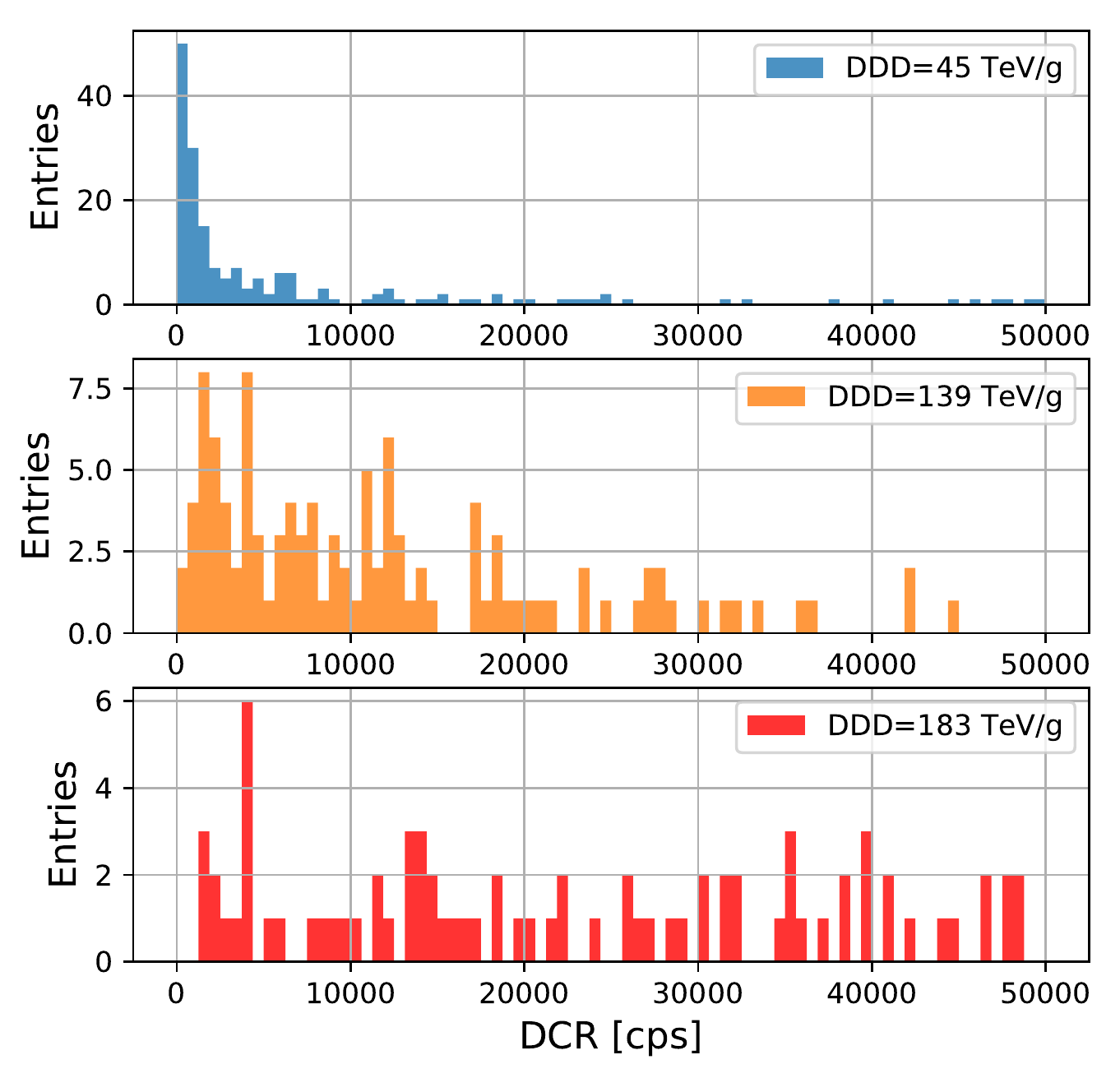}
  \caption{DCR distribution after proton irradiation at different displacement damage doses (DDD).}\label{population_after}
\end{figure}

This consideration is supported by the analysis of the DCR increase distribution at different DDD (Fig. \ref{population_after}). After the first step of the irradiation, the distribution presents a low-DCR peak followed by a long tail of hot SPADs.   As the DDD increases, the low-DCR peak is depopulated, while the events in the high-DCR tail increase. This behaviour is consistent with the effect of the typical displacement damage \cite{virmont}.
The reason is to be attributed to the small cross sections of the elastic and inelastic nuclear interactions, which are responsible for the displacement damage. 
As a consequence, the displacement damage occurrence at low fluence is small, and only a few SPADs are affected, precisely the ones in the high DCR tail.
For higher doses, the displacement damage probability per SPAD increases and a large part of the pixel are affected, resulting in a large non-uniformity increase and the creation of hot pixels.

\begin{figure}[b!]
\centering
  \includegraphics[width=8.5cm]{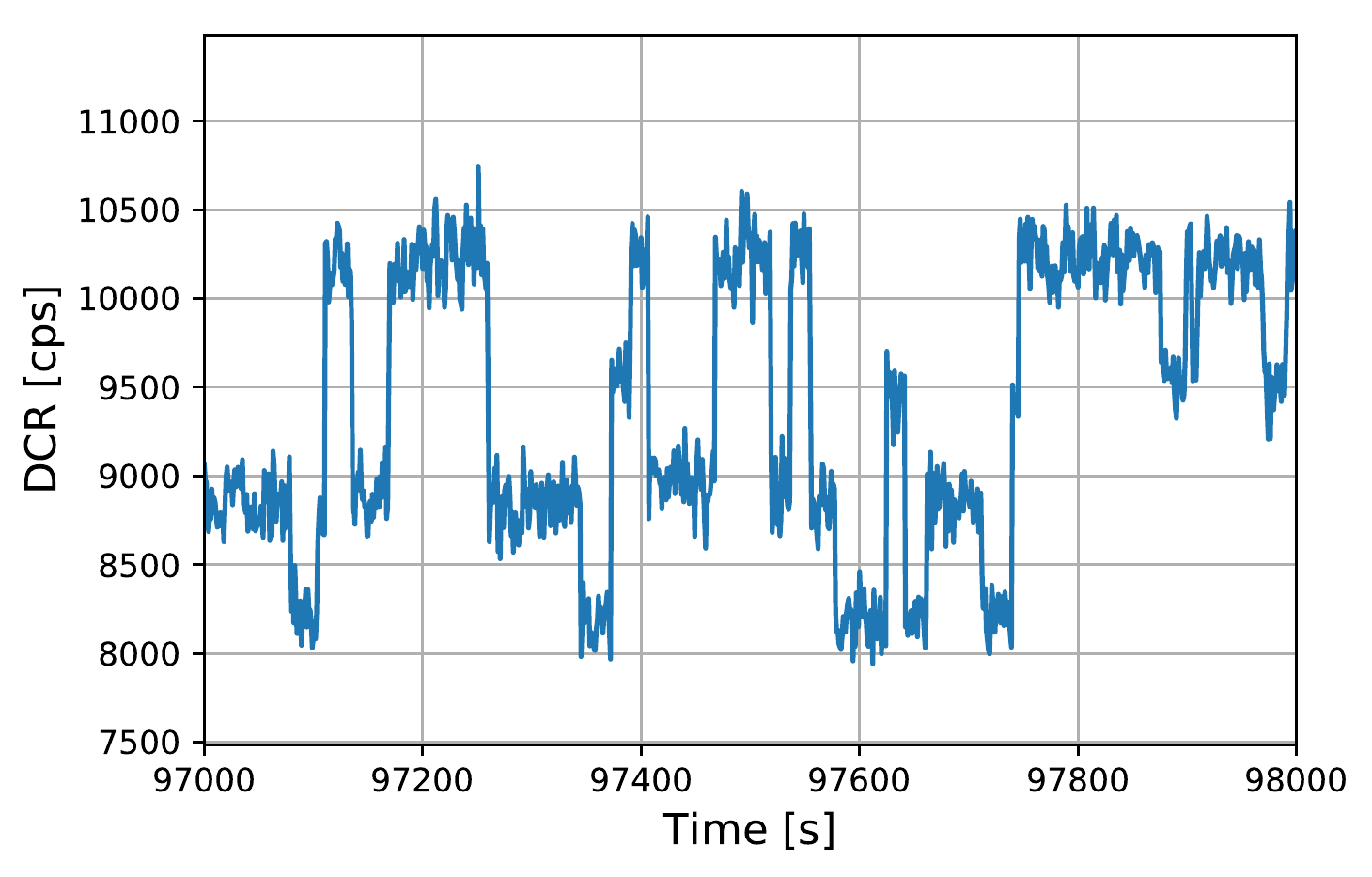}
\caption{DCR evolution with time of an irradiated SPADs exhibiting a dark count rate RTS behaviour. }\label{rts}

\end{figure}

Beyond the mean DCR increase, a further effect that we observed in many of the irradiated pixels is the appearance of a DCR Random Telegraph Signal (RTS) behaviour (Fig. \ref{rts}).  This consists of the discrete fluctuation of the DCR between two or more levels. RTS can also be attributed to the effect of displacement damage, i.e.,  to the reconfiguration of bulk defects \cite{DiCapua}.

\section{DCR increase mitigation}

One of the main advantages of CMOS SPADs compared to the analog ones is the possibility to address single cells. This allows one to disable high DCR pixels, resulting in a reducing  of the total DCR.
As an example, we report the case of a 5x5 SPADs matrix, irradiated at 94 TeV/g. By disabling the three noisiest cells, the average DCR reduces from 51405 cps to 17600 cps, meaning  a percentage decrease of 66\%.
This method can only be applied at relatively low doses, in fact, when the radiation levels are high, it would require to turn off too many pixels, causing a significant loss of efficiency.

\begin{figure}[b!]
\centering
  \includegraphics[width=8.5cm]{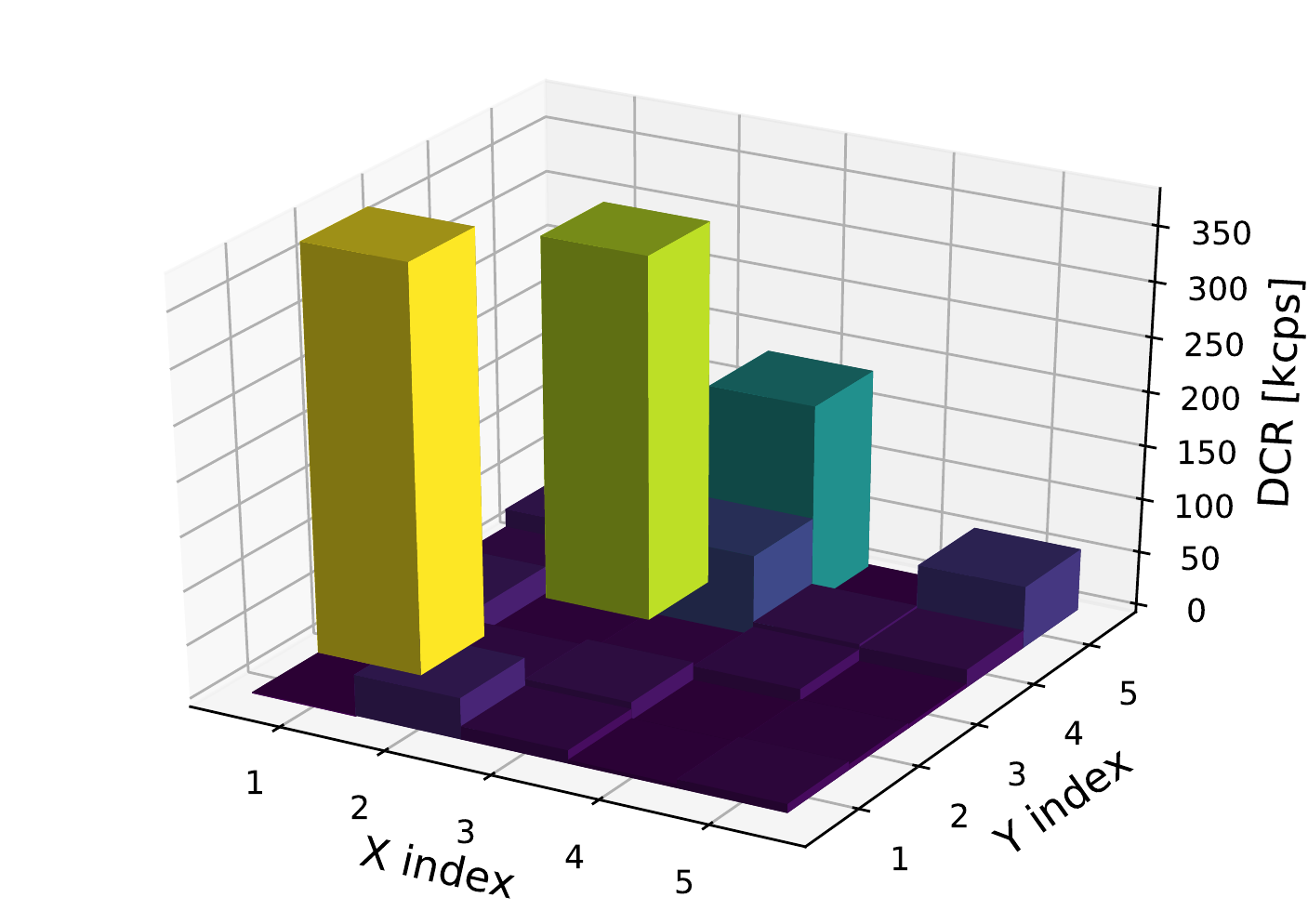}
  \caption{DCR maps of a 5x5 SPADs matrix irradiated at a DDD = 94 TeV/g.}\label{dcrddd_02}
\end{figure}

On the other hand, for high DCR levels, a coincidence between several pixels can significantly reduce the DCR. As an example, if we consider the worst DCR level obtained at a DDD = 608 TeV/g, a 3 SPADs coincidence would result in a few Hz false coincidence rate, assuming a time window of 10 ns.

\begin{figure}[ht]
\centering
  \includegraphics[width=8.5
 cm]{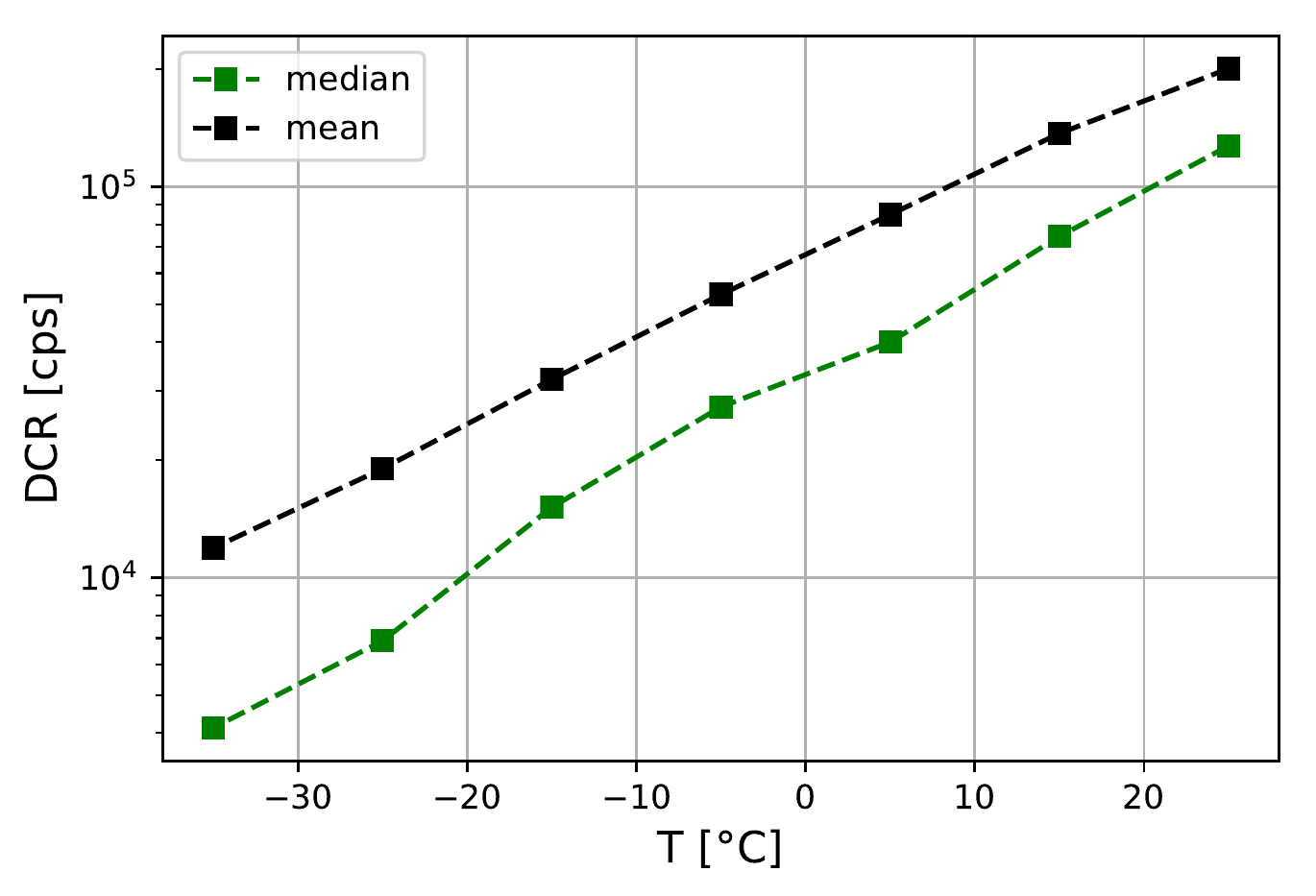}
  \caption{DCR as a function of the temperature step on proton irradiated SPADs (DDD = 608 TeV/g).}\label{cooling}
\end{figure}

Following, two further techniques to mitigate the DCR are reported, precisely cooling and annealing.
Using a climatic chamber we varied the temperature in the range [-35 $^\circ$C, +25 $^\circ$C]. Fig. \ref{cooling} shows the DCR as a function of temperature measured on SPADs irradiated at the maximum dose of 608 TeV/g.
As expected, the DCR  decreases  with temperature, roughly it halves its value by decreasing temperature every ten degrees. 
On a small subset SPADs, we measured the DCR activation energy  and found for the Pwell/Niso layout $E_a\sim 0.4eV$.
The activation energies are lower than the mid-bandgap, suggesting a generation mechanism related to the presence of bulk defects, with an electric field enhancement mechanisms, like Poole-Frenkel effects or Trap Assisted Tunneling (TAT) \cite{srour_enan, chantre}.

\begin{figure}[t]
\centering
  \includegraphics[width=8.5
 cm]{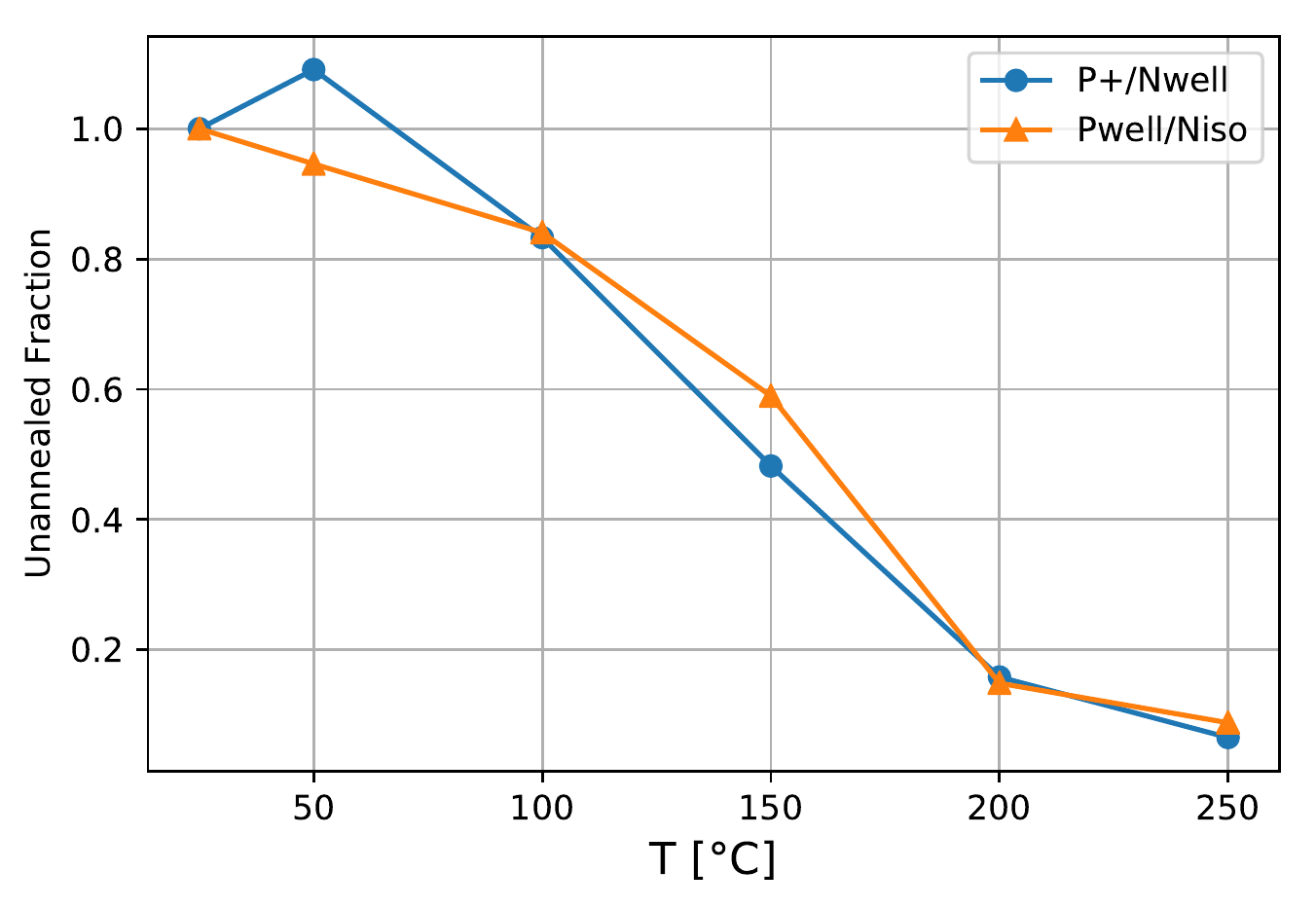}
  \caption{Evolution of unannealed fraction of DCR as a function of the annealing temperature step on proton irradiated SPADs at a DDD = 608 TeV/g.}\label{annealing}
\end{figure}

After the irradiation, samples have been left at room temperature (25 $^\circ$C) for a month. The subsequent measurements showed a small decrease (less than 5\%) in DCR for almost all SPADs, meaning that self-annealing effects occurred. 
In order to deepen this effect, we performed an high temperature isochronal annealing. 
Likewise to \cite{virmont}, we keep the samples for one hour at each temperature  between 50 $^\circ$C and 250 $^\circ$C with a 50-degree step. 
In Fig. \ref{annealing} we show the DCR unannealed factor for a chip irradiated at 608 TeV/g, namely the mean DCR measured at different annealing steps normalized to the initial value, i.e., before the first annealing step.
The annealing procedure is a useful tool for a deep study of the radiation induced defect types. Following \cite{Watkins,Nuns,Virmontois}, the detection of preferential annealing temperature allows to understand the type of defect involved in the DCR generation mechanisms. 
Most of DCR decrease in the curve (Fig. \ref{annealing}) is spread from 100 $^\circ$C up to 200 $^\circ$C. The non-sharp observed decrease, as pointed by \cite{Virmontois}, could be attributed to the combination of different types of defects with different annealing temperature or due to a strong contribution from cluster defects \cite{Srour}.
According to \cite{Watkins}, the defects corresponding to the annealing temperature interval reported in Fig. \ref{annealing} are dopant related vacancy complexes as V-P and V-As center defects. 
At the end of the annealing procedure, the average DCR for both SPAD layouts strongly decreases, and it becomes very close to the initial value measured before irradiation.


\section{Space Environment Simulation}
\begin{figure}[t]
\centering
  \includegraphics[width=8.5
 cm]{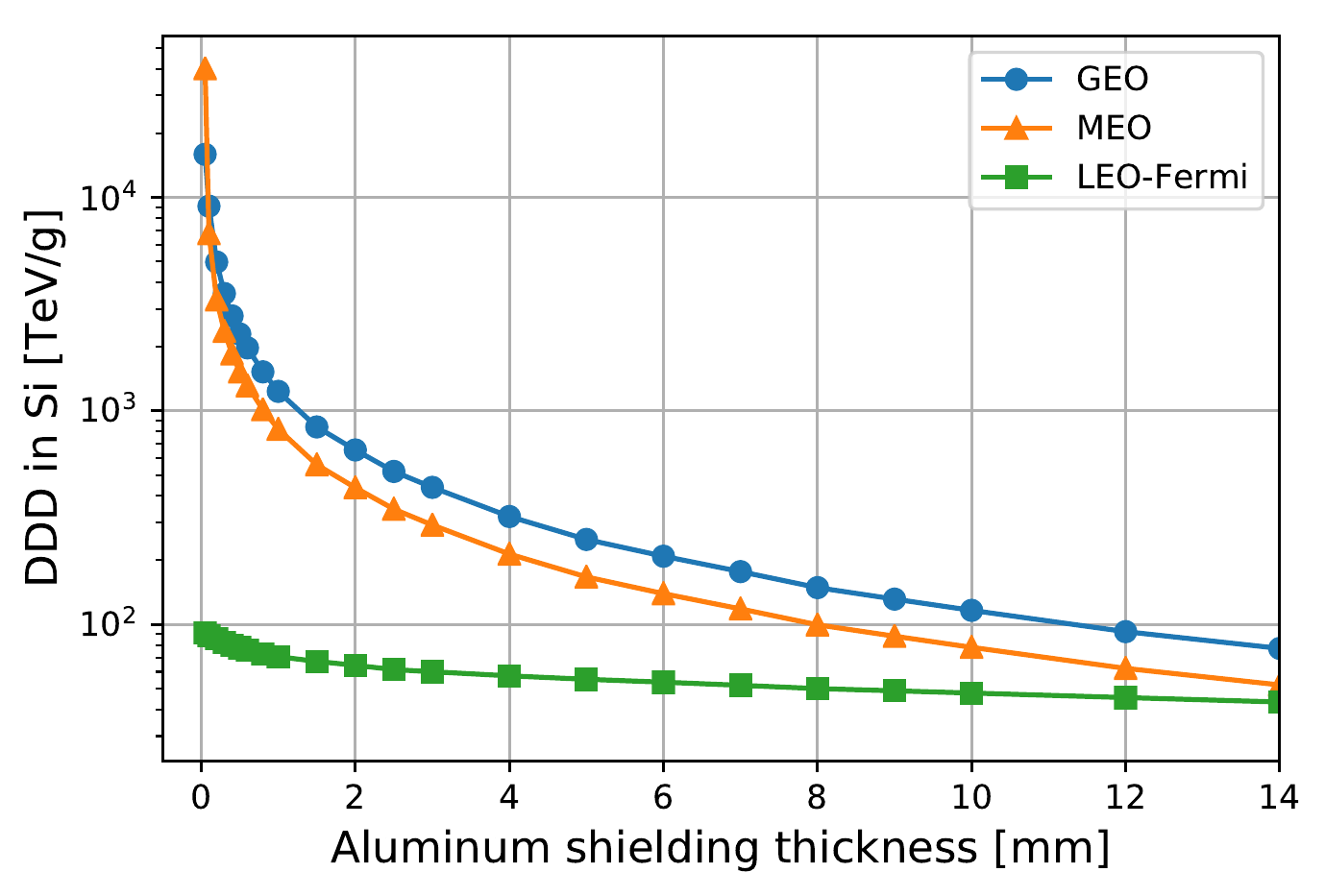}
  \caption{DDD in 10-year mission for typical LEO, MEO and GEO orbits (as specified in Table \ref{tab:simulations} ) as a function of the aluminum shielding thickness.}\label{space}
\end{figure}

In order to evaluate the applicability of CMOS SPADs in a space mission, we investigated the expected radiation doses for several space environment case-studies. As an example, we considered the Low-Earth Orbits (LEOs) followed by the Fermi Gamma-ray space-telescope and the International Space Station, a typical Medium Earth Orbit (MEO) and a Geostationary Orbit (GEO) exploited by many communication satellites. 
TID and DDD levels spread significantly by varying orbital altitude and inclination. LEO experiences space radiation mainly in the form of trapped protons and electrons within the Van Allen belts. Due to the much smaller radiation damage factor for electrons with respect to protons, the displacement damage term is dominated by protons. 
On the contrary, GEO does not experience DDD from trapped proton: the primary radiation sources are from solar protons and trapped electrons of the outer belt. 
The radiation dose was calculated using the online tools of the ESA Space Environment Information System (SPENVIS) \cite{Spenvis}. For Van Allen proton and electron radiation belts modeling, the AP8 \cite{Sawyer} and AE8 \cite{Vampola} models, calculated at the solar minimum, have been used. 
Fig. \ref{space} shows the expected DDD for typical LEO, MEO and GEO orbits as a function of aluminum shielding thickness for a 10-year mission. Table \ref{tab:simulations} summarizes the results of this analysis.

\begin{table*}[b]
\centering
    \caption{Expected ionizing and non-ionizing dose for different orbits, calculated for a 10-year mission lifetime and variable aluminum shielding. }
    \label{tab:simulations}
    \begin{tabular}{lccccc}
        \hline\hline
        Orbit Type & Mean Altitude & Inclination & Al Shield & TID & DDD \\
         & [km] & [deg.] & [mm] & [krad] & [TeV/g] \\
        \hline
        LEO-ISS & 400  & 51.6$^\circ$ & 2 & 3.7 & 19.6 \\
        LEO-FERMI & 560 & 26$^\circ$ & 2 & 3.0 & 64.3  \\
        MEO & 20000 & 56.2$^\circ$ & 3 & 2196 & 292 \\
        GEO & 36000 & 0$^\circ$ & 3 & 355 & 438  \\
        \hline\hline
    \end{tabular}
\end{table*}

On the basis of the previously showed DCR increase analysis and the space environments simulation, we can draw some conclusions about the applicability of the devices in space. Low orbits can be considered safe enough for long lifetime space missions: only a slight DCR increase is expected. On the contrary, results showed a significant increase in DCR at a proton fluence corresponding to long lifetime, high altitude missions. This could be an issue for using SPADs in most of the applications in long life GEO mission, requiring the implementation of an efficient mitigation procedure.

\section{Conclusion}
The purpose of this work is the understanding and the modelling of the effects of radiation on the dark count rate in CMOS SPADs. This study has been done in order to check the suitability of CMOS SPADs for future applications requiring single-photon detection in radiation environments, like space or HEP collider experiments. For this purpose, a proton irradiation campaign has been conducted on many samples of a test chip containing two SPADs structures. Samples were irradiated at several doses, demonstrating a strong susceptibility to the radiation damage. 
We observed that DCR increase is proportional to the amount of energy loss for displacement damage by the incoming particle. 
The observed DCR increase can be a critical performance issue for CMOS SPADs that are used in a radiation environment. We investigated cooling and annealing, as possible methods of mitigation. 
After isochronal annealing, the DUTs showed a good recovery of the original DCR behaviour.  
We also investigated the expected displacement damage doses for several space mission case-studies, and found that CMOS SPADs can be safely employed in low-inclination LEO space missions.





\bibliography{sample}

\begin{thebibliography}{10}
\expandafter\ifx\csname url\endcsname\relax
  \def\url#1{\texttt{#1}}\fi
\expandafter\ifx\csname urlprefix\endcsname\relax\def\urlprefix{URL }\fi
\expandafter\ifx\csname href\endcsname\relax
  \def\href#1#2{#2} \def\path#1{#1}\fi

\bibitem{Cova}
S.~Cova, A.~Longoni, A.~Andreoni, {Towards picosecond resolution with
  single-photon avalanche diodes}, Review of Scientific Instruments 52 (1981)
  408 -- 412.
\newblock \href {http://dx.doi.org/10.1063/1.1136594}
  {\path{doi:10.1063/1.1136594}}.

\bibitem{Rochas}
A.~{Rochas}, A.~R. {Pauchard}, P.~. {Besse}, D.~{Pantic}, Z.~{Prijic}, R.~S.
  {Popovic}, Low-noise silicon avalanche photodiodes fabricated in conventional
  cmos technologies, IEEE Transactions on Electron Devices 49~(3) (2002)
  387--394.
\newblock \href {http://dx.doi.org/10.1109/16.987107}
  {\path{doi:10.1109/16.987107}}.

\bibitem{Palubiak}
D.~P. {Palubiak}, M.~J. {Deen}, Cmos spads: Design issues and research
  challenges for detectors, circuits, and arrays, IEEE Journal of Selected
  Topics in Quantum Electronics 20~(6) (2014) 409--426.
\newblock \href {http://dx.doi.org/10.1109/JSTQE.2014.2344034}
  {\path{doi:10.1109/JSTQE.2014.2344034}}.

\bibitem{Vilella}
E.~Vilella, O.~Alonso, A.~Diéguez, {3D integration of Geiger-mode avalanche
  photodiodes aimed to very high fill-factor pixels for future linear
  colliders}, Nucl. Instrum. Meth. A731 (2013) 103--108.
\newblock \href {http://dx.doi.org/10.1016/j.nima.2013.05.022}
  {\path{doi:10.1016/j.nima.2013.05.022}}.

\bibitem{Apix}
L.~Pancheri, A.~Ficorella, P.~Brogi, G.~Collazuol, G.-F. Dalla~Betta,
  P.~S.~Marrocchesi, F.~Morsani, L.~Ratti, A.~Savoy-Navarro, A.~Sulaj, First
  demonstration of a two-tier pixelated avalanche sensor for charged particle
  detection, IEEE Journal of the Electron Devices Society 5 (2017) 404--410.
\newblock \href {http://dx.doi.org/10.1109/JEDS.2017.2737778}
  {\path{doi:10.1109/JEDS.2017.2737778}}.

\bibitem{farich}
D.~A.~Finogeev, A.~B.~Kurepin, V.~I.~Razin, A.~I.~Reshetin, E.~Usenko,
  A.~Barnyakov, M.~Yu.~Barnyakov, V.~S.~Bobrovnikov, A.~Buzykaev,
  P.~V.~Kasyanenko, S.~Kononov, E.~Kravchenko, I.~Kuyanov, A.~P.~Onuchin,
  I.~Ovtin, N.~A.~Podgornov, A.~A.~Talyshev, A.~Danilyuk, Development of farich
  detector for particle identification system at accelerators, Physics of
  Particles and Nuclei 49 (2018) 30--32.
\newblock \href {http://dx.doi.org/10.1134/S1063779618010100}
  {\path{doi:10.1134/S1063779618010100}}.

\bibitem{tof}
M.~Bohm, A.~Lehmann, S.~Motz, F.~Uhlig, Fast sipm readout of the panda tof
  detector, Journal of Instrumentation 11 (2016) C05018--C05018.
\newblock \href {http://dx.doi.org/10.1088/1748-0221/11/05/C05018}
  {\path{doi:10.1088/1748-0221/11/05/C05018}}.

\bibitem{dsipm}
Y.~Hsmisch, T.~Frach, C.~Degenhardt, A.~Thon, Fully digital arrays of silicon
  photomultipliers (dsipm) – a scalable alternative to vacuum photomultiplier
  tubes (pmt), Physics Procedia 37 (2012) 1546--1560.
\newblock \href {http://dx.doi.org/10.1016/j.phpro.2012.03.749}
  {\path{doi:10.1016/j.phpro.2012.03.749}}.

\bibitem{jinstspad}
Z.~Liu, M.~Pizzichemi, E.~Auffray, P.~Lecoq, M.~Paganoni, Performance study of
  philips digital silicon photomultiplier coupled to scintillating crystals,
  Journal of Instrumentation 11 (2016) 1017--1017.
\newblock \href {http://dx.doi.org/10.1088/1748-0221/11/01/P01017}
  {\path{doi:10.1088/1748-0221/11/01/P01017}}.

\bibitem{Shockley}
W.~Shockley, W.~T.~J.~Read, Statistics of the recombination of holes and
  electrons, Physical Review - PHYS REV X 87 (1952) 835--842.
\newblock \href {http://dx.doi.org/10.1103/PhysRev.87.835}
  {\path{doi:10.1103/PhysRev.87.835}}.

\bibitem{Hall}
R.~N.~Hall, Electron-hole recombination in germanium, Physical Review - PHYS
  REV X 87 (1952) 387--387.
\newblock \href {http://dx.doi.org/10.1103/PhysRev.87.387}
  {\path{doi:10.1103/PhysRev.87.387}}.

\bibitem{Kane}
E.~O.~Kane, Theory of tunneling, Journal of Applied Physics 32 (1961) 83 -- 91.
\newblock \href {http://dx.doi.org/10.1063/1.1735965}
  {\path{doi:10.1063/1.1735965}}.

\bibitem{Xu}
Y.~Xu, P.~Xiang, X.~Xie, Comprehensive understanding of dark count mechanisms
  of single-photon avalanche diodes fabricated in deep sub-micron cmos
  technologies, Solid-State Electronics 129 (2017) 168 -- 174.
\newblock \href {http://dx.doi.org/10.1016/j.sse.2016.11.009}
  {\path{doi:10.1016/j.sse.2016.11.009}}.

\bibitem{Srour}
J.~R. {Srour}, J.~W. {Palko}, Displacement damage effects in irradiated
  semiconductor devices, IEEE Transactions on Nuclear Science 60~(3) (2013)
  1740--1766.
\newblock \href {http://dx.doi.org/10.1109/TNS.2013.2261316}
  {\path{doi:10.1109/TNS.2013.2261316}}.

\bibitem{Li}
Z.~Li, Y.~Xu, C.~Liu, Y.~Gu, F.~Xie, Y.~Li, H.~Hu, X.~Zhou, X.~Lu, X.~Li,
  S.~Zhang, Z.~Chang, J.~Zhang, Z.~Xu, Y.~Zhang, J.~Zhao, Characterization of
  radiation damage caused by 23 mev protons in multi-pixel photon counter
  (mppc), Nuclear Instruments and Methods in Physics Research Section A
  Accelerators Spectrometers Detectors and Associated Equipment 822 (2016)
  63--70.

\bibitem{Andreotti}
M.~Andreotti, W.~Baldini, R.~Calabrese, G.~Cibinetto, A.~Cotta~Ramusino,
  C.~De~Donato, R.~Faccini, M.~Fiorini, E.~Luppi, R.~Malaguti, A.~Montanari,
  A.~Pietropaolo, V.~Santoro, G.~Tellarini, L.~Tomassetti, N.~Tosi, Study of
  the radiation damage of silicon photo-multipliers at the gelina facility,
  Journal of Instrumentation 9 (2014) P04004.
\newblock \href {http://dx.doi.org/10.1088/1748-0221/9/04/P04004}
  {\path{doi:10.1088/1748-0221/9/04/P04004}}.

\bibitem{Moscatelli}
F.~Moscatelli, M.~Marisaldi, P.~Maccagnani, C.~Labanti, F.~Fuschino, M.~Prest,
  A.~Berra, D.~Bolognini, M.~Ghioni, I.~Rech, A.~Gulinatti, A.~Giudice,
  G.~Simmerle, A.~Candelori, S.~Mattiazzo, X.~Sun, J.~Cavanaugh, D.~Rubini,
  Radiation tests of single photon avalanche diode for space applications,
  Nuclear Instruments and Methods in Physics Research Section A: Accelerators,
  Spectrometers, Detectors and Associated Equipment 711 (2013) 65–72.
\newblock \href {http://dx.doi.org/10.1016/j.nima.2013.01.056}
  {\path{doi:10.1016/j.nima.2013.01.056}}.

\bibitem{Charbon}
Y.~Li, C.~Veerappan, M.-J. Lee, W.~Lin, Q.~Guo, E.~Charbon, A
  radiation-tolerant, high performance spad for sipms implemented in cmos
  technology, Proceedings of 10th International Workshop on Radiation Effects
  on Semiconductor Devices for Space Applications (2016) 1--4.\href
  {http://dx.doi.org/10.1109/NSSMIC.2016.8069762}
  {\path{doi:10.1109/NSSMIC.2016.8069762}}.

\bibitem{Charbon2}
M.~Gersbach, C.~Niclass, L.~Carrara, M.~Sergio, N.~Scheidegger, H.~Shea,
  E.~Charbon, A study of the effects of gamma radiation on cmos single-photon
  avalanche diodes, Proceedings of 7th IEEE Sensors Conference (2008) 919--921.

\bibitem{Ratti}
L.~Ratti, P.~Brogi, G.~Collazuol, G.-F. Dalla~Betta, A.~Ficorella, L.~Lodola,
  P.~Marrocchesi, S.~Mattiazzo, F.~Morsani, M.~Musacci, L.~Pancheri, C.~Vacchi,
  Dark count rate degradation in cmos spads exposed to x-rays and neutrons,
  IEEE Transactions on Nuclear Science 66 (2019) 567--574.
\newblock \href {http://dx.doi.org/10.1109/TNS.2019.2893233}
  {\path{doi:10.1109/TNS.2019.2893233}}.

\bibitem{campajola_2019}
M.~Campajola, F.~D. Capua, D.~Fiore, C.~Nappi, E.~Sarnelli, L.~Gasperini,
  Radiation effects on single-photon avalanche diodes manufactured in deep
  submicron {CMOS} technology, Journal of Physics: Conference Series 1226
  (2019) 012007.
\newblock \href {http://dx.doi.org/10.1088/1742-6596/1226/1/012007}
  {\path{doi:10.1088/1742-6596/1226/1/012007}}.

\bibitem{philips}
M.~Yu.~Barnyakov, T.~Frach, S.~Kononov, I.~Kuyanov, V.~Prisekin, Radiation
  hardness test of the philips digital photon counter with proton beam, Nuclear
  Instruments and Methods in Physics Research Section A Accelerators
  Spectrometers Detectors and Associated Equipment 824 (2016) 83 -- 84.
\newblock \href {http://dx.doi.org/10.1016/j.nima.2015.10.098}
  {\path{doi:10.1016/j.nima.2015.10.098}}.

\bibitem{Karami}
M.~A. Karami, A.~Pil-ali, S.~M.~R. Safaee, Multistable defect characterization
  in proton irradiated single-photon avalanche diodes, Optical and Quantum
  Electronics 47 (2014) 1--6.
\newblock \href {http://dx.doi.org/10.1007/s11082-014-0089-7}
  {\path{doi:10.1007/s11082-014-0089-7}}.

\bibitem{Karami2}
A.~Karami, L.~Carrara, C.~Niclass, M.~Fishburn, E.~Charbon, Rts noise
  characterization in single-photon avalanche diodes, Electron Device Letters,
  IEEE 31 (2010) 692 -- 694.
\newblock \href {http://dx.doi.org/10.1109/LED.2010.2047234}
  {\path{doi:10.1109/LED.2010.2047234}}.

\bibitem{DiCapua}
F.~Di~Capua, M.~Campajola, L.~Campajola, C.~Nappi, E.~Sarnelli, L.~Gasparini,
  H.~Xu, Random telegraph signal in proton irradiated single-photon avalanche
  diodes, IEEE Transactions on Nuclear Science 65 (2018) 1654--1660.
\newblock \href {http://dx.doi.org/10.1109/TNS.2018.2814823}
  {\path{doi:10.1109/TNS.2018.2814823}}.

\bibitem{Spenvis}
Esa space environment information system (ese),
  \url{http://www.spenvis.oma.be/}.

\bibitem{FBK}
Fondazione bruno kessler, \url{https://www.fbk.eu/en/}.

\bibitem{Pancheri}
L.~Pancheri, D.~Stoppa, Low-noise single photon avalanche diodes in 0.15 $\mu$m
  cmos technology, Proceedings of European Solid-State Device Research
  Conference (2011) 179 -- 182\href
  {http://dx.doi.org/10.1109/ESSDERC.2011.6044205}
  {\path{doi:10.1109/ESSDERC.2011.6044205}}.

\bibitem{Hesong}
H.~Xu, L.~Pancheri, L.~Braga, G.-F. Dalla~Betta, D.~Stoppa, Cross-talk
  characterization of dense single-photon avalanche diode arrays in cmos 150-nm
  technology, Optical Engineering 55 (2016) 067102.
\newblock \href {http://dx.doi.org/10.1117/1.OE.55.6.067102}
  {\path{doi:10.1117/1.OE.55.6.067102}}.

\bibitem{pancheri_field}
L.~Pancheri, D.~Stoppa, G.-F.~D. Betta, Characterization and modeling of
  breakdown probability in sub-micrometer cmos spads, IEEE Journal of Selected
  Topics in Quantum Electronics 20 (2014) 328--335.

\bibitem{srim}
J.~F. Ziegler, M.~Ziegler, J.~Biersack, Srim – the stopping and range of ions
  in matter (2010), Nuclear Instruments and Methods in Physics Research Section
  B: Beam Interactions with Materials and Atoms 268~(11) (2010) 1818 -- 1823,
  19th International Conference on Ion Beam Analysis.
\newblock \href {http://dx.doi.org/https://doi.org/10.1016/j.nimb.2010.02.091}
  {\path{doi:https://doi.org/10.1016/j.nimb.2010.02.091}}.

\bibitem{fluka}
T.~Böhlen, F.~Cerutti, M.~Chin, A.~Fassò, A.~Ferrari, P.~Ortega, A.~Mairani,
  P.~Sala, G.~Smirnov, V.~Vlachoudis, The fluka code: Developments and
  challenges for high energy and medical applications, Nuclear Data Sheets 120
  (2014) 211 -- 214.
\newblock \href {http://dx.doi.org/https://doi.org/10.1016/j.nds.2014.07.049}
  {\path{doi:https://doi.org/10.1016/j.nds.2014.07.049}}.

\bibitem{NIEL}
Niel calculator, \url{http://www.sr-niel.org/}.

\bibitem{trantid}
M.~{Manghisoni}, L.~{Ratti}, V.~{Re}, V.~{Speziali}, Radiation hardness
  perspectives for the design of analog detector readout circuits in the
  0.18-/spl mu/m cmos generation, IEEE Transactions on Nuclear Science 49~(6)
  (2002) 2902--2909.
\newblock \href {http://dx.doi.org/10.1109/TNS.2002.805413}
  {\path{doi:10.1109/TNS.2002.805413}}.

\bibitem{tranddd}
G.~C. {Messenger}, A summary review of displacement damage from high energy
  radiation in silicon semiconductors and semiconductor devices, IEEE
  Transactions on Nuclear Science 39~(3) (1992) 468--473.
\newblock \href {http://dx.doi.org/10.1109/23.277547}
  {\path{doi:10.1109/23.277547}}.

\bibitem{virmont}
C.~{Virmontois}, V.~{Goiffon}, P.~{Magnan}, S.~{Girard}, C.~{Inguimbert},
  S.~{Petit}, G.~{Rolland}, O.~{Saint-Pe}, Displacement damage effects due to
  neutron and proton irradiations on cmos image sensors manufactured in deep
  submicron technology, IEEE Transactions on Nuclear Science 57~(6) (2010)
  3101--3108.
\newblock \href {http://dx.doi.org/10.1109/TNS.2010.2085448}
  {\path{doi:10.1109/TNS.2010.2085448}}.

\bibitem{srour_enan}
J.~R. {Srour}, R.~A. {Hartmann}, Enhanced displacement damage effectiveness in
  irradiated silicon devices, IEEE Transactions on Nuclear Science 36~(6)
  (1989) 1825--1830.
\newblock \href {http://dx.doi.org/10.1109/23.45375}
  {\path{doi:10.1109/23.45375}}.

\bibitem{chantre}
G.~Vincent, A.~Chantre, D.~Bois, Electric field effect on the thermal emission
  of traps in semiconductor junctions, Journal of Applied Physics 50 (1979)
  5484 -- 5487.
\newblock \href {http://dx.doi.org/10.1063/1.326601}
  {\path{doi:10.1063/1.326601}}.

\bibitem{Watkins}
G.~Watkins, Intrinsic defects in silicon, Materials Science in Semiconductor
  Processing 3 (2000) 227--235.
\newblock \href {http://dx.doi.org/10.1016/S1369-8001(00)00037-8}
  {\path{doi:10.1016/S1369-8001(00)00037-8}}.

\bibitem{Nuns}
T.~Nuns, G.~Quadri, J.~David, O.~Gilard, Annealing of proton-induced random
  telegraph signal in ccds, Nuclear Science, IEEE Transactions on 54 (2007)
  1120 -- 1128.
\newblock \href {http://dx.doi.org/10.1109/TNS.2007.902351}
  {\path{doi:10.1109/TNS.2007.902351}}.

\bibitem{Virmontois}
C.~Virmontois, V.~Goiffon, P.~Magnan, O.~Saint-Pe, S.~Girard, S.~Petit,
  G.~Rolland, A.~Bardoux, Total ionizing dose versus displacement damage dose
  induced dark current random telegraph signal in cmos image sensors, IEEE
  Transactions on Nuclear Science - IEEE TRANS NUCL SCI 58 (2011) 3085--3094.
\newblock \href {http://dx.doi.org/10.1109/TNS.2011.2171005}
  {\path{doi:10.1109/TNS.2011.2171005}}.

\bibitem{Sawyer}
D.~M.~Sawyer, J.~I.~Vette, Ap8 trapped proton environment for solar maximum and
  solar minimum, NASA STIRecon Technical Report N 77 (1977) 18983.

\bibitem{Vampola}
A.~L. {Vampola}, Outer zone energetic electron environment update, in:
  Conference on the High Energy Radiation Background in Space. Workshop Record,
  1997, pp. 128--136.
\newblock \href {http://dx.doi.org/10.1109/CHERBS.1997.660263}
  {\path{doi:10.1109/CHERBS.1997.660263}}.

\end{thebibliography}

\end{document}